\documentclass[reprint,superscriptaddress]{revtex4-1}
\usepackage{amsmath,graphicx}
\usepackage{dsfont}
\usepackage{textcomp}
\usepackage{comment}

\begin{document}

\title{Wavelength sensitivity of the speckle patterns produced by an integrating sphere}

\author{Morgan Facchin} \email{mf225@st-andrews.ac.uk}
\affiliation{SUPA School of Physics and Astronomy, University of St Andrews, North Haugh, St Andrews KY16 9SS, UK}

\author{Kishan Dholakia}
\affiliation{SUPA School of Physics and Astronomy, University of St Andrews, North Haugh, St Andrews KY16 9SS, UK}
\affiliation{Department of Physics, College of Science, Yonsei University, Seoul 03722, South Korea}
\affiliation{Graduate School of Engineering, Chiba University, 1-33 Yayoi-cho, Inage-ku, Chiba-shi 263-0022, japan}

\author{Graham D. Bruce}
\affiliation{SUPA School of Physics and Astronomy, University of St Andrews, North Haugh, St Andrews KY16 9SS, UK}

\begin{abstract}
Speckle metrology is a powerful tool in the measurement of wavelength and spectra. Recently, speckle produced by multiple reflections inside an integrating sphere has been proposed and showed high performance. However, to our knowledge, a complete characterisation of speckle sensitivity to wavelength in that geometry has not been performed to date.
In this work, we derive a general model predicting the variation in a speckle pattern as a result of a generic transformation. 
Applying this to a shift in the incident wavelength, we show that the speckle sensitivity is mainly governed by the radius and surface reflectivity of the sphere. 
We show that integrating spheres offer sensitivity three orders of magnitude above that of multimode fibres of a similar size, and discuss analogies with the transmission line of a Fabry-Pérot interferometer.
\end{abstract}
\maketitle

\section{Introduction}

Speckle patterns are granular intensity patterns that are the result of the interference of coherent light reflecting off a rough surface. Despite their random nature, they are rich in information and can be sensitive to various effects, which make them an interesting tool for metrology. Among many applications, we find the measurement of displacement \cite{archbold70,wang2006core,wang2006vortex,wang2005phase}, vibrations \cite{bianchi14}, polarisation \cite{facchin2020pol}, blood flow in tissues \cite{Briers13}, speech and heartbeat \cite{zalevsky09}, and drying processes in paint \cite{van2016paint}.

The applications of speckle metrology that we focus upon here particularly are recently identified topics, namely spectrometry \cite{Wan15,redding2013all,Redding12,Redding14,Liew16,Redding13,wan2020high} and measurements of wavelength \cite{wan2020high,chakrabarti2015speckle,bruce19,Bruce20,Mazilu14,odonnellhigh,Gupta19,Metzger17,davila2020single}. Both rely on the sensitivity of speckle patterns to a change in incident laser wavelength. For the remainder of this work, we define sensitivity as the HWHM (half width at half maximum) of some measure of change in the speckle pattern as a function of wavelength change, which is also the commonly used definition for the resolution of speckle spectrometers \cite{Cao17}. This sensitivity naturally depends on the way the speckles are produced. The most common methods for producing speckles are reflection on a rough surface \cite{chakrabarti2015speckle}, propagation through a multimode fibre \cite{Wan15,redding2013all,Redding12,Redding14,Liew16,bruce19,Bruce20}, or through a disordered medium \cite{Redding13,Gupta19,Mazilu14}. 

In this work we focus on an alternative way of producing speckle patterns, which is due to multiple reflections of light inside an integrating sphere \cite{odonnellhigh,Gupta19,Metzger17,davila2020single,Boreman90}. This has proven to produce sensitive speckles with an intensity distribution close to a gamma distribution. To the best of our knowledge, in such a geometry, a rigorous theoretical understanding of the sensitivity in terms of key experimental parameters is lacking. This would allow a comparison to be made between media generating speckle patterns, to determine any trade offs and to make an informed choice between these various schemes. To this end we derive a general model predicting the change occurring in a speckle pattern, first for a generic transformation, then specifically for a wavelength change. We find that the key parameters are the sphere's radius and its surface reflectivity. We then compare this to the sensitivity obtained for the case of a speckle pattern produced by a multimode fibre, and discuss an interesting analogy with Fabry-Pérot interferometers.

\section{Derivation of the speckle similarity profile}
\begin{figure}[h!] 
\centering\includegraphics[width=8cm]{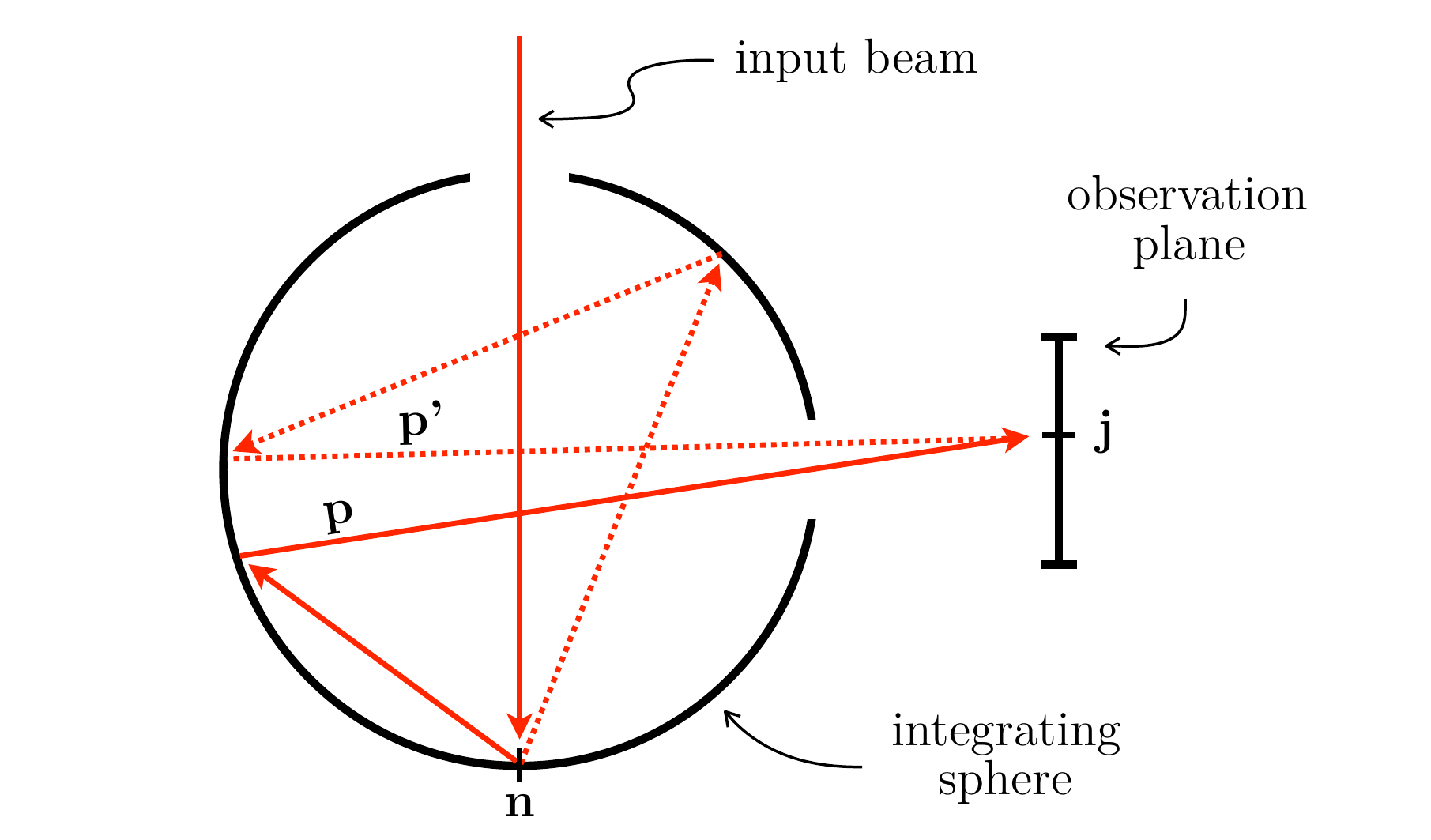}
\caption{Geometry of the problem. The input beam is incident upon the inner surface of a spherical cavity and undergoes multiple diffusive reflections. The light that escapes the sphere forms a speckle pattern which is collected at the observation plane. $n$ is an element of the inner surface, $j$ is a point of the observation plane, and $p$ and $p'$ are two possible paths from $n$ to $j$. }
\label{fig:geometry}
\end{figure}

The problem is as follows: a beam of monochromatic light enters a spherical cavity of radius $R$ and uniform reflectivity $\rho$, as shown in Fig. \ref{fig:geometry}. We consider that the inner surface presents a Lambertian reflectance, which implies that the surface is rough, and therefore the light exiting the sphere forms a speckle pattern. Now we apply an arbitrary transformation to the system (which could be a change in wavelength, refractive index of the medium, or a deformation of the sphere) and we ask how the speckle changes as a consequence. 

Denoting $I$ and $I'$ as the intensity patterns before and after the transformation respectively, we quantify the change in the speckle by the following quantity:

\begin{equation} \label{eq:correl}
\begin{split}
S&= \Big\langle  \Big( \frac{I_{j}-\langle I_{j}\rangle_j}{\sigma} \Big)\Big( \frac{I_{j}'-\langle I_{j}'\rangle_j}{\sigma'} \Big) \Big\rangle _j,  \\
\end{split}
\end{equation}

\noindent where $I_{j}$ is the intensity observed at point $j$ of the observation plane, $\sigma$ is the standard deviation of the intensity pattern, and $\langle \rangle_j$ denotes averaging over the observation plane.
The quantity $S$ is called similarity (or Pearson correlation coefficient), and quantifies the morphological change between the two images. It leads to a value of 1 for identical speckle patterns and decreases towards 0 as they diverge from one another. Here we seek an expression for the similarity as a function of relevant parameters of the sphere and the applied transformation, and apply it to the case of wavelength change. 

We assume that the coherence length of the light is large compared to the spread of the path length distribution in the sphere (equal to $4R/(3 \ln{\rho})$ \cite{hodgkinson2009using}), so that the light can be considered fully coherent on the observation plane. The intensity at point $j$ then verifies  

\begin{equation} \label{eq:intensity}
I_{j}\propto\left | E_j \right |^2=E_{j}^\dagger E_{j},
\end{equation}

\noindent with $E_{j}$ the $3\times1$ complex field at $j$, and $\dagger$ denoting the conjugate transpose. We work in the monochromatic approximation, where the time dependence disappears after multiplication by the conjugate. Therefore we omit the time dependence as well as the time averaging. 

It can be shown \cite[p.~41]{Goodman} that the similarity of the absolute square of the field is equal to the absolute square of the field's correlation: 

\begin{equation} \label{mu}
    S=\left |  
    \frac{\big\langle E_j^\dagger E_j'\big\rangle_j}
    {\big\langle \left | E_j \right |^2 \big\rangle_j} 
    \right |^2,
\end{equation}

\noindent where we assume that the statistics of the speckle pattern are the same before and after transformation, namely $\langle \left | E_j' \right |^2 \rangle_j=\langle \left | E_j \right |^2 \rangle_j=\langle I_j \rangle_j$.

Now we can develop $E_j$ by modelling the journey of the light between the illuminated region and the observation plane. The inner surface is modelled by an assembly of $M$ discrete surface elements, with $M$ large enough for each element to be considered flat. The field at $j$ can then be written as the sum of the contributions from each surface element illuminated by the input beam, reading ${E}_{j}=\sum_{n}{E}_{nj}$, 
with ${E}_{nj}$ the contribution of surface element $n$ to the field at $j$. Furthermore, the field diffuses everywhere in space from $n$ to $j$, and ${E}_{nj}$ implicitly contains the contributions of all the possible paths starting from $n$ and ending at $j$. As a transformation affects each path differently (in the general case), let us decompose ${E}_{nj}$ into the contributions of all paths, reading ${E}_{nj}=\sum_{p}\alpha_{njp}{E}_{n}$, 
where we label each path by an index $p$. ${E}_{n}$ is the field coming from the incident beam at $n$, and $\alpha_{njp}$ is a $3\times3$ complex matrix describing the change in the field from $n$ to $j$ following path $p$. The number of paths between any $n$ and $j$ is of course infinite (one can think for example of an arbitrary long alternation between the two same elements). These two decompositions follow from the superposition principle and the assumption that the diffusion is linear, and together give us an expression for ${E}_{j}$:

\begin{equation}
    E_j=\sum_{np}\alpha_{njp}{E}_{n}.
\end{equation}

Inserting this in (\ref{mu}), and using the distribution property of the conjugate transpose, we have 

\begin{equation} \label{mudev}
\begin{split}
     S&=\left |  
    \frac{\Big\langle {E}_{n}^\dagger\alpha_{njp}^\dagger \alpha_{njp}'{E}_{n}\Big\rangle_{njp} }
    {\Big\langle  {E}_{n}^\dagger\alpha_{njp}^\dagger \alpha_{njp}{E}_{n}  \Big\rangle_{njp}} 
    \right |^2,
\end{split}
\end{equation}

\noindent where we assumed that the fields coming from different paths or different elements of the illuminated region are uncorrelated. 

The matrix $\alpha_{njp}$ can be decomposed into the product of an amplitude, phase, and polarisation term, reading $\alpha_{njp}=\sqrt{T_{njp}} e^{i\varphi_{njp}}U_{njp}$, where $T_{njp}$ is the intensity transmission of path $p$ from $n$ to $j$, $\varphi_{njp}$ is the phase acquired by the field along the path, and $U_{njp}$ is a $3\times3$ unitary matrix changing the polarisation. 
Furthermore, we are interested here in transformations that leave the incident beam profile unchanged, therefore $E_n$ is constant and the effect of a transformation appears in $\alpha_{njp}$ only. This effect typically appears in the phase term, so that we can write $\alpha_{njp}'=\alpha_{njp} e^{i\phi_{njp}}$, with $\phi_{njp}=\varphi_{njp}'-\varphi_{njp}$ the phase shift induced by the transformation. We use the symbol $\phi_{njp}$ instead of $\Delta\varphi_{njp}$ to avoid heavy notations, as only $\phi_{njp}$ appears in the following.

Inserting these expressions for $\alpha_{njp}$ and $\alpha_{njp}'$ in (\ref{mudev}), and using the orthogonality property of unitary matrices ($U^\dagger U=I$ with $I$ the identity matrix), we find 

\begin{equation} 
      S=\left |  
    \frac{\Big\langle T_{njp} \left | {E}_{n} \right |^2 e^{i\phi_{njp}}  \Big\rangle_{njp} }
    {\Big\langle  T_{njp} \left | {E}_{n} \right |^2  \Big\rangle_{njp}} 
    \right |^2.
\end{equation}

This can be simplified if we choose the size of our surface elements to be large compared to the small scale asperities of the inner surface, in which case $T_{njp}$ earns the macroscopic properties of the Lambertian reflectance, and loses its $n$ dependence. Its $j$ dependence can be neglected in any case, as all paths impinge on $j$ from almost identical angles and distances. We also neglect the $n$ and $j$ dependence of $\phi_{njp}$, as they only account for small contributions at the endpoints of the path. These simplifications lead to 

\begin{equation} \label{Tp}
\begin{split}
S =  \left |  \frac{   \sum_{p}   T_{p} e^{i \phi_p}  }
{\sum_{p} T_{p}       } \right |^2.
\end{split}
\end{equation}

\begin{figure}[h!] 
\centering\includegraphics[width=\linewidth]{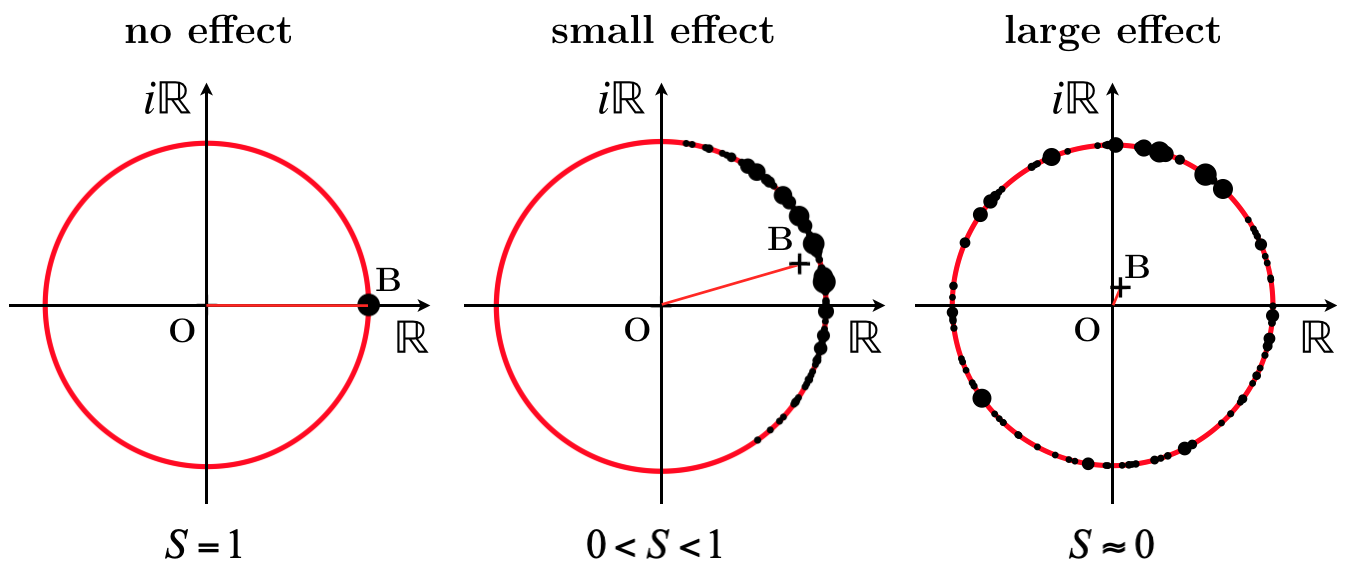}
\caption{Visual representation of equation (\ref{Tp}). Each path is represented by a black dot, which together form an infinite cluster on the unit circle of the complex plane. The size and azimuthal position represent respectively the transmission of the path ($T_p$) and the phase shift induced by the transformation on that path ($\phi_p$). The similarity (S) is equal to the square of the distance between the origin (O) and the barycentre of the cluster (B). We show three stages where the effect of the transformation increases from none to high.  }
\label{fig:cluster}
\end{figure}

We can recognise in (\ref{Tp}) a weighted average of the phase factors, where the weights are given by the transmission of the paths. This lends itself to a nice visual interpretation in the complex plane (see Fig. \ref{fig:cluster}). Plotting each phase factor as a point in the complex plane (each corresponding to a path) forms an infinite cluster lying on the unit circle. The similarity is the square of the distance between the barycentre of this cluster and the origin. When no transformation is applied ($\phi_p=0$ for all paths), all the points are at 1 and the similarity is therefore 1 (no speckle change). As the effect of a transformation increases, the points spread out on the unit circle and the barycentre approaches the origin (hence a decreasing value of the similarity) until the points are uniformly spread, where the similarity is close to zero. 

Now $T_p$ can be shown to be a simple function of the number of single passes that make up path $p$ (we define single pass as a straight line linking two surface elements of the sphere, a path is a succession of single passes). Indeed, the Lambertian reflectance, combined with the spherical geometry, conspire to make the transmission a constant for each single pass. This can be shown in the following way. Consider one surface element emitting a power $P$ in the volume of the sphere, and another element receiving some part of this power. The Lambertian reflectance implies that the received power is $P'= P \rho \, \delta S \cos{\theta_1}\cos{\theta_2}/(\pi d^2)$ \cite{carr1997integrating}, with $d$ the distance between the elements, $\theta_1$ and $\theta_2$ the angles between their normal and the line joining them, and $\delta S$ their area. Now the spherical geometry imposes a relation between these quantities, namely $d=2R\cos{\theta}$, with $\theta=\theta_1=\theta_2$. If we insert this in the expression of the received power (and recall that $\delta S=4\pi R^2/M$) we find that it simplifies to $P\rho/M$. Therefore, the transmission is $\rho/M$ for each single pass, and the transmission of a full path made of $N$ single passes is $T_p=(\rho/M)^{N(p)}$, which is a great simplification of the problem.

With this in mind, we can split the sums in (\ref{Tp}) into groups of paths that are made of the same number $N$ of single passes

\begin{equation} \label{integral}
\begin{split}
S &=  \left |   \sum_{N} \sum_{p'}   \frac{\rho^N}{\sum_{N}\rho^N}  \frac{e^{i \phi_{p'}}}{M^N}  \right |^2 \\
 &\approx \left | \int_0^{\infty} \frac{\rho^N}{-\ln{\rho}} \: e^{iN\mu -N\sigma^2/2} \; dN\right |^2, 
\end{split}
\end{equation}

\noindent where $p'$ designates the paths that are made of $N$ single passes. The first line makes appear the average value of $e^{i \phi_{p'}}$ in the population $p'$ (as $M^N$ also happens to be the total number of paths made of $N$ single passes), which can be expressed in more explicit terms. Indeed, the phase $\phi_{p'}$ acquired on a full path is the sum of the phases acquired on each successive single path, so that we can approximate $\phi_{p'}$ by a Gaussian random variable, as a consequence of the central limit theorem. Besides, statistics tell us that the complex exponential of a Gaussian random variable $G(\mu,\sigma)$ has average $\langle e^{iG(\mu,\sigma)} \rangle=e^{i\mu -\sigma^2/2}$. Therefore, we have $\sum _{p'} e^{i \phi_{p'}}/M^N =e^{iN\mu -N\sigma^2/2}$, with $\mu$ the average phase induced by the transformation on a single pass, and $\sigma$ the standard deviation of this phase. In the second line we approximate the sum by an integral, which has the advantage of giving a simpler form and impacts little the accuracy of the result. 
We set the lower limit to $N=0$, again for simplicity of the final form. This choice is not critical as, when $\rho$ approaches unity (typically a sphere will have $\rho>0.9$), more power goes to higher values of $N$ and the choice of the starting point does not influence the outcome to any appreciable degree. 
Performing the integral finally gives

\begin{equation} \label{Sgeneral}
S =   \frac{ 1 }
{\big (1-\frac{\sigma^2}{2\ln{\rho}} \big)^2+\big (\frac{\mu}{\ln{\rho}} \big)^2     }.
\end{equation}

This expression is valid for any transformation that applies a phase shift of average $\mu$ and standard deviation $\sigma$ to the field along one single pass. It is interesting to note in passing that any effect for which the $\mu$ term dominates leads to a Lorentzian profile, while any effect where the $\sigma$ term dominates leads to the square of a Lorentzian (if we recall that $\ln{\rho}$ is negative).

We collect here the assumptions made in our model: the incident light is monochromatic, with a coherence length large compared to the spread of the path length distribution, the inner surface has a Lambertian reflectance with uniform reflectivity, and the diffusion is linear. The last steps between (\ref{integral}) and (\ref{Sgeneral}) also assume a high reflectivity ($\rho > 0.9$).

\section{Sensitivity to wavelength variations}
Let us now apply (\ref{Sgeneral}) to the case of a wavelength variation. When light propagates along a path of length $z$, it acquires a spatial phase $kz$, with $k$ the wavenumber. When the wavelength changes, it induces a phase change on the path equal to $\Delta k \,z$. Here we see that the effect of the transformation indeed takes the form of a phase factor which is different for each path, with $\phi_p=\Delta k \, z_p$. It follows that the average phase change on a single pass is $\mu= \Delta k z_0$, with $z_0$ the average distance between two points in a sphere, that is, the average chord length. This is given by geometry to be $4R/3$ \cite{Fry:06,berengut1972random,sidiropoulos2014n}. Likewise, the standard deviation of chord length is $\sqrt{2}R/3$ \cite{berengut1972random,sidiropoulos2014n}. This gives 


\begin{equation} \label{stat}
\mu=\frac{4}{3}R\Delta k \quad \quad \quad \sigma=\frac{\sqrt{2}}{3}R\Delta k.
\end{equation}


Inserting these expressions in (\ref{Sgeneral}), it can be shown that we are in a case where the $\mu$ term dominates, and therefore the similarity shows a Lorentzian profile:

\begin{equation} \label{eq:lorentzian_wav}
S=  \frac{1}{1+\left(\frac{ \Delta \lambda}{ \Delta \lambda_0}\right)^2 },
\end{equation} 
with $\Delta \lambda_0=3\lambda^2\left|\ln\rho\right|/(8\pi R)$, which also corresponds to the HWHM of the Lorentzian. For modest parameters such as $R=1$ cm, $\rho=$ 0.9, and $\lambda=$ 780 nm, this gives already a fairly high sensitivity with an HWHM of about 0.8 pm. 

We note that, for a wavemeter, the smallest change in wavelength that can be measured is much smaller than this HWHM, and will rather correspond to the smallest change in $S$ which is detectable over sources of experimental noise. For example, attometre-resolved measurements were realised with an MMF where the HWHM was 620 pm \cite{bruce19}.

\section{Experimental verification}
\begin{figure}[h!] 
\centering\includegraphics[width=\linewidth]{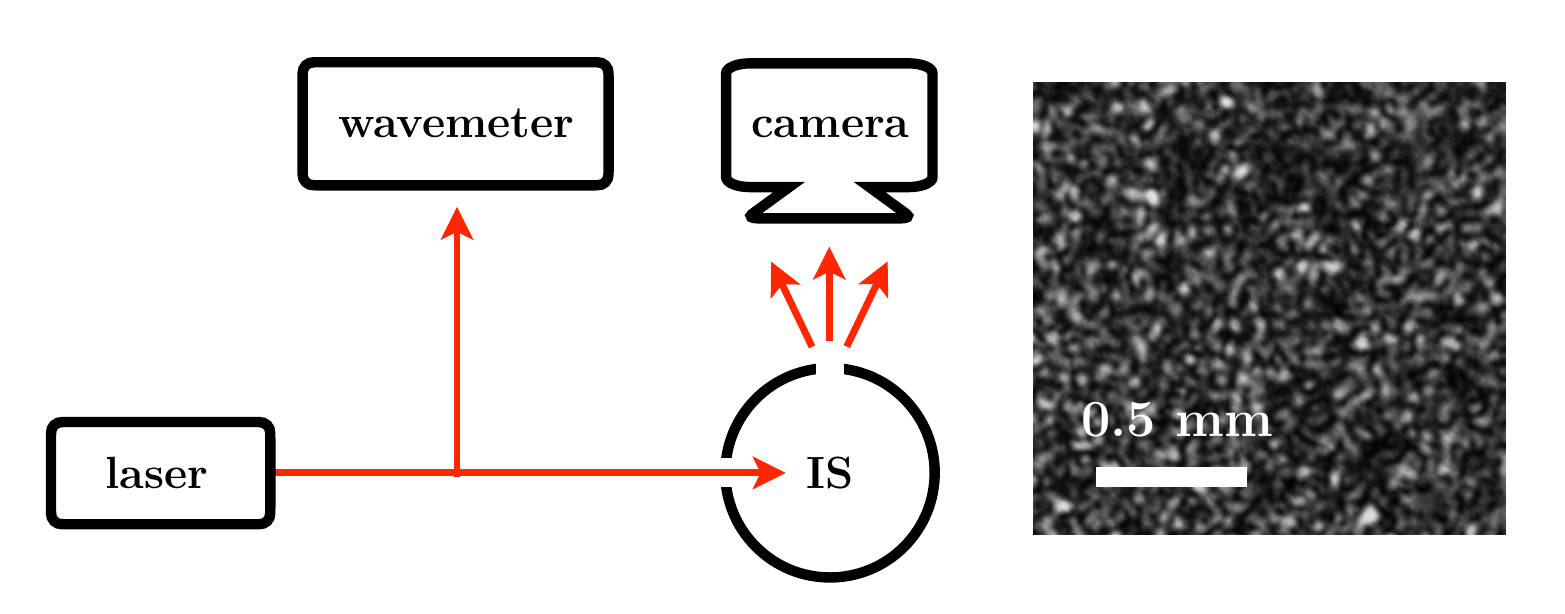}
\caption{Experimental setup. Laser light enters an integrating sphere (IS) and produces a speckle pattern collected on a camera. The wavelength of the laser is then changed in a linear manner, while the resulting speckle change is recorded. The wavelength change is monitored by a reference wavemeter. An example of a $200\! \times \! 200$-pixel speckle pattern image is shown.  }
\label{fig:setup}
\end{figure}

In order to verify (\ref{eq:lorentzian_wav}), we implement the experimental setup shown in Fig. \ref{fig:setup}. A laser beam (of 780 nm wavelength, 10 mW power, and having a coherence length of a few km (Toptica DLPro)) is injected in an integrating sphere, and the resulting speckle pattern is recorded on a CMOS camera (Mikrotron MotionBLITZ EoSens mini2). We place the camera so that the individual speckle grains cover hundreds of pixels, in order to minimise spatial averaging effects. Larger images minimise the variance of the similarity across different realisations, we use $200\! \times \! 200$-pixels images (about a few hundred speckle grains) which offers a good compromise between variance and computation speed. We use a 1.25 cm radius integrating sphere, carved into a 3 cm edge aluminium cube and manually coated with Spectraflect. The light enters and escapes the sphere via two 3 mm diameter holes.

We then apply a linear wavelength variation by applying a triangular modulation to tune the cavity length of the laser. The amplitude of the wavelength variation is $3.1 \pm 0.05$ pm and is measured using a Fizeau-based wavemeter (HighFinesse WS7). 

\begin{figure}[h!] 
\centering\includegraphics[width=\linewidth]{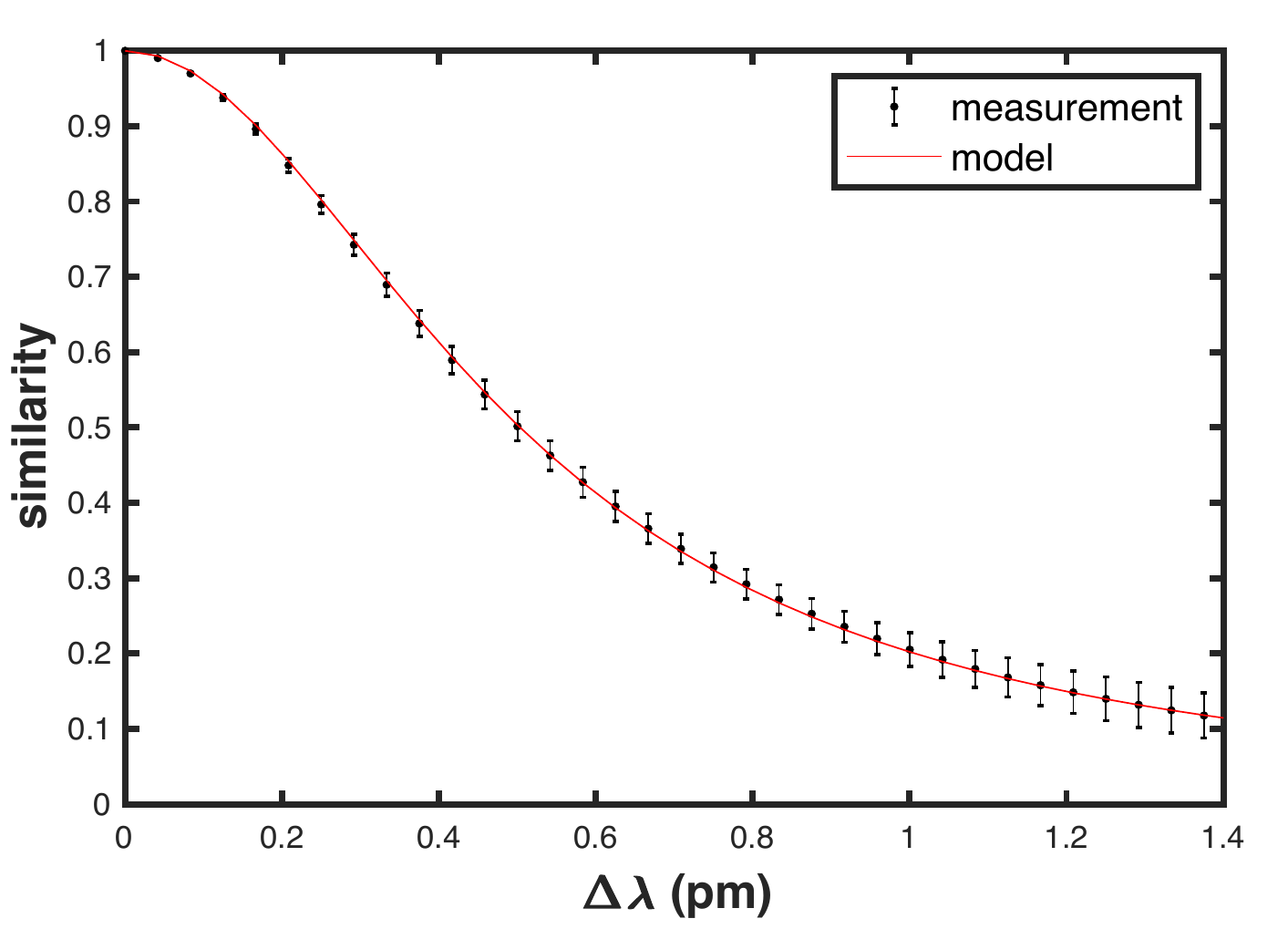}
\caption{Similarity as a function of wavelength change, experimental (black) and Lorentzian profile predicted by model (red), fitted for a reflectivity $\rho=0.917$. The centre and span of the error bars respectively give the mean and standard deviation of the set of curves extracted from the data set. The HWHM is 0.5 pm.}
\label{fig:simil_wav}
\end{figure}

One similarity profile can be extracted by computing the similarity between one reference image and the rest of the data set. By using several reference images across the data set, we extract several similarity profiles. We show the average and standard deviation in Fig. \ref{fig:simil_wav}. We fit the resulting profile using (\ref{eq:lorentzian_wav}) with $\rho$ as a free parameter, as it is the most uncertain quantity. We find a good agreement for $\rho=0.917 \pm 0.002$. The uncertainty comes in equal amount from that of the wavelength modulation amplitude and the fitting.

The fit in Fig. \ref{fig:simil_wav} confirms the predicted functional form, though we recognise that we have $\rho$ as a free parameter. In order to perform an independent measurement of $\rho$, we use a method based on the measurement of the output power \cite{carr1997integrating}. The power measured at a certain distance from the output port can be expressed as a function of the input power, port size, port-detector distance, detector diameter, integrating sphere diameter, and reflectivity. We measure the output power at different distances from the output port using a photodiode power sensor (Thorlabs S121C) and extract an estimation of the reflectivity $\rho=0.908 \pm 0.009$. Here the main sources of uncertainty are the machine precision limit on the output port diameter (3\%) and the uncertainty of the power meter measurement (7\%). The two values of $\rho$ agree within one $\sigma$ of uncertainty. Other sources of systematic error that are difficult to assess were not taken into account, such as the alignment of the detector with the port and orientation of the detector (which both lead to an underestimation of $\rho$). Another source of systematic error could be possible non-uniformity of the reflectivity across the inner surface.  

We note in passing that the fit of the similarity curve can serve as a means of measuring the reflectivity, with an accuracy only limited by the knowledge of the applied wavelength variation and the sphere's radius.

\section{Comparison with speckle patterns generated by multimode fibres}
In this section we compare the sensitivity of an integrating sphere to that of a multimode fibre. The similarity profile of a multimode fibre is not Lorentzian, but we know the dependence of its HWHM on the relevant fibre parameters, namely $\Delta \lambda_0 \propto \lambda^2/(L\, N\!A^2)$, with $\lambda$ the wavelength, $L$ the fibre's length, and $N\!A$ its numerical aperture \cite{Rawson80,redding2013all}. It is independent of the core size above a critical diameter of about 100 $\mu$m \cite{redding2013all}. For a step-index fibre, the relationship was empirically found to be $\Delta \lambda_0 \approx 2.4 \, \lambda^2/(L\, N\!A^2)$ \cite{redding2012using}. Equating this HWHM to that found above for the integrating sphere, we find a direct proportionality between the fibre's length and the radius of the equivalent integrating sphere. With a standard value $N\!A=0.22$, and our previously found reflectivity $\rho=0.917$, we have 

\begin{equation}
    L \approx 7000\,R,
\end{equation}

\noindent which means that a sphere of radius $R$ shows the same sensitivity as a $7000R$-long fibre. For example, our sphere of radius 1.25 cm is equivalent to a fibre of length 90 m. 

This demonstrates that an integrating sphere can offer a very compact alternative to an optical fibre, as the effective space occupied by a fibre is much larger than that of the equivalent integrating sphere (even though its intrinsic volume is smaller) for similar performance. 
Another advantage of the sphere is that the sensitivity to wavelength change is independent of the way in which light is coupled into the sphere. In contrast, it was shown that the sensitivity of multimode fibres depends strongly on the number of spatial modes excited in the fibre and therefore on the coupling of light at the fibre input \cite{velsink2021comparison,redding2013all}. Moreover, an integrating sphere offers the additional advantage of being more robust to mechanical perturbations, as they are monolithic and have no moving parts, which can be a serious difficulty when using fibres.

\section{Comparison with the spectral linewidth of a Fabry-Pérot interferometer} \label{perot}
Interestingly, the similarity profile (\ref{eq:lorentzian_wav}) has the same functional form as the transmission line of a Fabry-Pérot interferometer. In fact, this is not so surprising as one could tackle this problem using an approach similar to our model, which would give the same expressions (\ref{Tp}) and (\ref{Sgeneral}), where instead $S$ would be the output intensity normalised to maximum. 
For a Fabry-Pérot, the HWHM is $\lambda^2\ln{\rho}/(4 \pi L)$ \cite{Ismail:16}, with $L$ the distance between the two mirrors, and $\rho$ their reflectivity. 

For sake of comparison, let us consider a sphere and a Fabry-Pérot of the same reflectivity, with the sphere's diameter equal to the length of the Fabry-Pérot ($L=2R$). In these conditions we have that the HWHM of the Fabry-Pérot line is exactly 3 times smaller than that of the sphere's similarity. 

This can be understood qualitatively, as the length of a Fabry-Pérot (L) is larger than the average length in the sphere ($4R/3$), the latter being exactly 3/2 smaller. From this simple observation, however, we would expect the HWHM of the Fabry-Pérot to be 3/2 times smaller than that of the sphere, not 3. The additional factor 2 comes from the one-dimensional flavour of the Fabry-Pérot. Indeed, any increase in length in the Fabry-Pérot must come in multiples of 2L, not L. Therefore the average length of a single pass (which is actually a round trip) is $2L$, and the substitution $2L \Leftrightarrow 4R/3$ is what allows the correct translation between the two cases. Interestingly, the lower dimensionality of the Fabry-Pérot system is in fact beneficial for sensitivity.  

Of course, the HWHM is not the only parameter of interest for a wavemeter or spectrometer, but also the bandwidth or range over which the wavelength measurement can be performed. For a Fabry-Perot cavity, the wavelength is retrieved modulo $\Delta \lambda_{FSR}$ (with $\Delta \lambda_{FSR} = \lambda^2 / L$ in air). However, the higher-dimensional nature of speckle removes this degeneracy: any two wavelengths separated by more than a few HWHMs are essentially orthogonal, and the range over which the wavemeter operates is in principle only limited by the size of the calibration set. In practice, this is usually limited by the spectral window of the camera \cite{Metzger17}, although the limit may be further reduced by finite sampling of the speckle \cite{Redding13}.

\section{Summary and conclusion}
We have derived a general expression for the change occurring in the speckle pattern produced by an integrating sphere resulting from a generic transformation. The amount of change is quantified by the similarity (\ref{eq:correl}), for which we give an explicit expression (\ref{Sgeneral}). This expression depends only on the average and standard deviation of the phase shift induced by the transformation on a single pass through the sphere.

In the case of wavelength variation, the similarity becomes a simple Lorentzian profile (\ref{eq:lorentzian_wav}), whose HWHM depends mainly on the surface reflectivity and the radius of the sphere. We tested this result experimentally and found good agreement. The measurement of this Lorentzian profile can be used as an accurate, easy-to-implement means of measuring integrating spheres' reflectivity that is free from systematic error. By comparing this to the speckle pattern produced by transmission through a multimode fibre, we showed that an integrating sphere of radius $R$ gives the same sensitivity to wavelength change as a fibre of length $\approx 7000R$, with standard parameters.

The sphere's similarity profile has the same functional form as the transmission line of a Fabry-Pérot interferometer. For a sphere and a Fabry-Pérot of the same reflectivity, with the sphere's diameter equal to the length of the Fabry-Pérot ($L=2R$), we found that the HWHM of the Fabry-Pérot line is exactly 3 times smaller than that of the sphere's similarity. 

This work suggests that the importance of the integrating sphere in the context of wavelength measurement has been overlooked, offering significant advantages when compared to the alternative methods discussed here. The model developed here can be adapted to consider other effects which transform the speckle, and will enable the optimised design of integrating spheres for speckle metrology.

\section{Acknowledgements}
This work was supported by funding from the Leverhulme Trust (RPG‐2017‐197) and the UK Engineering and Physical Sciences Research Council (EP/P030017/1).

\bibliography{sample}
\end{document}